\documentstyle[11pt,amssymb,amsfonts,amsthm,amssymb,amscd,amstext,epsfig]{article}

\def\ben{\begin{equation}}
\def\een{\end{equation}}

\let\a=\alpha \let\b=\beta \let\g=\gamma \let\d=\delta 
  \let\q=\theta 
     \let\r=\rho

\let\w=\omega  \let\D=\Delta  
 \let\P=\Phi

\def\be{\begin{equation}}
\def\ee{\end{equation}}
\def\ba{\begin{array}}
\def\ea{\end{array}}

\def\vp{\varphi}

\def\dalemb#1#2{{\vbox{\hrule height .#2pt
        \hbox{\vrule width.#2pt height#1pt \kern#1pt
                \vrule width.#2pt}
        \hrule height.#2pt}}}

\newcommand{\bea}{\begin{eqnarray}}
\newcommand{\eea}{\end{eqnarray}}

\def\rs{r_{*}}

\def\R{{{\Bbb R}}}

\begin{document}
\begin{flushright}
\hfill{DAMTP-2003-42} \\
{hep-th/0305001}
\end{flushright}

\begin{center}
\vspace{1cm} { \Large {\bf Instability of generalised AdS black holes
and thermal field theory}}

\vspace{1.5cm}

Sean A. Hartnoll

\vspace{0.3cm}

s.a.hartnoll@damtp.cam.ac.uk

\vspace{0.8cm}

{\it DAMTP, Centre for Mathematical Sciences,
 Cambridge University\\ Wilberforce Road, Cambridge CB3 OWA, UK}

\vspace{2cm}

\end{center}

\begin{abstract}

We study black holes in AdS-like spacetimes, with the horizon
given by an arbitrary positive curvature Einstein metric. A
criterion for classical instability of such black holes is found
in the large and small black hole limits. Examples of large
unstable black holes have a B\"ohm metric as the horizon. These,
classically unstable, large black holes are locally
thermodynamically stable. The gravitational instability has a dual
description, for example by using the $AdS_7 \times S^4$
version of the AdS/CFT correspondence. The instability
corresponds to a critical temperature of the dual thermal field
theory defined on a curved background.

\end{abstract}

\pagebreak
\setcounter{page}{1}

\section{Introduction}

Gravitational physics in higher dimensions, and black hole physics
in particular, allows a wealth of phenomena that are not possible
in the more tightly constrained dynamics of four dimensional
spacetime. This paper will study one such phenomenon in the
context of the correspondence between gravity in Anti-de Sitter
(AdS) spacetime and gauge theories in one less dimension
\cite{Maldacena:1997re,Gubser:1998bc,Witten:1998qj}. Black holes
play an important r\^{o}le in understanding this correspondence at
finite temperature \cite{Witten:1998zw}. Anti-de Sitter black hole
spacetimes contain information about the thermodynamics of a dual
field theory. For standard AdS black holes, the dual field theory
is defined on the background $S^1\times S^d$. For the generalised black
holes to be introduced shortly, the field theory is instead on the
background $S^1\times B^{\prime}$, with $B^{\prime}$ an arbitrary
Einstein manifold.

We will exhibit an instability of generalised black holes with a
negative cosmological constant and predict a corresponding effect
in the dual thermal field theory. The most immediate embedding of
these results into a known duality turns out to be into the
correspondence relating M-theory on $AdS_7
\times S^4$ and the six dimensional conformal field theory
describing M5 branes at low energy \cite{Maldacena:1997re}.
The paper has three parts:
\begin{itemize}
  \item Derivation of the criterion for classical instability of
  generalised black holes with a negative cosmological constant (\S 2 - \S 4).
  \item Discussion of the relationship between classical instability and
  thermodynamic stability of these black holes (\S 5).
  \item Discussion of the dual description of the instability in
  thermal field theory (\S 6).
\end{itemize}
Let us briefly motivate these three points in turn.

One important feature of higher dimensional gravity is that there
are many possibilities for constructing non-asymptotically flat
spacetimes. Asymptotic flatness itself in higher dimensions appears
to be more subtle than in four dimensions \cite{Hollands:2003ie}.
The natural $D$-dimensional generalisation of the four dimensional
Schwarzschild black hole, with vanishing cosmological constant,
was written down many years ago \cite{Tangherlini}. One may then
prove that this is the unique regular static black hole that is
asymptotically flat \cite{Hwang, Gibbons:2002av}. However, if one
drops the requirement of asymptotic flatness, then the Einstein
equations allow the replacement of the sphere that forms the
horizon with any positive curvature Einstein manifold, $B$,
appropriately normalised. This $B$ is related to the $B^{\prime}$
on which the dual field theory is defined by a rescaling of the
metric. Even if one wishes to retain the spherical topology of the
horizon, a countable infinity of such metrics are known on spheres
of dimension $d=5\ldots9$, constructed by B\"ohm \cite{bohm}. With a
negative cosmological constant, other possibilities exist, in
which the horizon metric may have negative or zero curvature
\cite{Birmingham:1998nr, Vanzo:1997gw, Brill:1997mf, Mann:1996gj},
but these will not be considered here.

A natural question concerns the stability of these new black hole
spacetimes. Classical instabilities of generalised black holes in
flat space, i.e. with a vanishing cosmological constant, were
studied in some detail in \cite{Gibbons:2002pq}. A criterion for
instability was found that depended on the lowest Lichnerowicz
eigenvalue of the horizon manifold, $B$. If the eigenvalue is less
than a critical value, the spacetime is unstable.

The unstable mode in question is a transverse tracefree tensor
harmonic of the horizon metric, other modes are not expected to be
dangerous \cite{Gibbons:2002pq}. Such modes do not exist in four
spacetime dimensions, because the horizon $S^2$ does not admit
nonzero tensor harmonics \cite{Higuchi:1986wu}, so the instability
is inherently higher dimensional. In seven spacetime dimensions,
examples of stable flat space black holes are given when the
horizon manifold is an Einstein-Sasaki manifold, such as $T^{1,1}$
or $S^5$ \cite{Gibbons:2002th}. Examples of unstable black holes
in this dimension are provided when the horizon is a B\"ohm metric
on $S^5$ or $S^2\times S^3$ \cite{Gibbons:2002th}. The possibility
of both stable and unstable generalised black holes shows that a
na\"{\i}ve global thermodynamic argument \cite{Gibbons:2002av},
that any positive curvature horizon must have volume and hence
entropy lower than a spherical horizon \cite{Bishop}, is
insufficient to understand classical instability. For example,
some of the stable Einstein-Sasaki metrics on $S^2\times S^3$ have
much lower volume than the unstable B\"ohm metrics on the same
topology \cite{Bergman:2001qi,Gibbons:2002th}.

This paper considers the classical stability of generalised black
holes in Anti-de Sitter space, that is, with a negative
cosmological constant. In \S 2 the formalism is set up and the
perturbation mode is described. In \S 3 an instability criterion
is derived in the large black hole limit. The dependence on the
horizon size is found analytically whilst the dependence on the
dimension is found numerically. Larger black holes are more likely
to be stable. In particular, for a fixed horizon manifold, there
is always a critical size above which black holes are stable. On
the other hand, for a fixed, arbitrarily large horizon size, there
exists a B\"ohm metric that one may use as the horizon manifold,
such that the resulting black hole is unstable. In
\S 4 small black holes are discussed and it is shown that the
criterion for instability of small generalised black holes is the
same as the criterion in flat space, as one might expect.

Recent work on large charged black holes in Anti-de Sitter space
lead to the conjecture by Gubser and Mitra that translationally
invariant black branes are classically stable if and only if they
are locally thermodynamically stable
\cite{Gubser:2000mm,Gubser:2000ec}. This conjecture has since been
strengthened and generalised \cite{Reall:2001ag,Hubeny:2002xn}.
Local thermodynamic stability is the statement that the Hessian of
the entropy with respect to extensive quantities is negative
definite. In the uncharged case that we are interested in, this
reduces to the requirement of positive specific heat. We see in \S
5 that the unstable large black holes are locally
thermodynamically stable. This result is discussed in the context
of the Gubser-Mitra conjecture. We conclude that, unlike spherical
black holes, the large generalised black hole limit cannot be
thought of as a translationally invariant black brane limit, and
therefore the conjecture is not contradicted. However, the
situation suggests that one has to be cautious about using
thermodynamic arguments to explain classical instabilities.

Finally, in \S 6 we discuss the dual field theory implications of
the instability. The geometry of the generalised black holes at a
fixed radius is $S^1 \times B^{\prime}$, where $B^{\prime}$ is
just $B$ rescaled by a factor of the radius squared. In
particular, this is true at large radius and is therefore the
background on which the dual thermal field theory is defined.
The $S^1$ is of course the time direction. We will see that the
instability is naturally translated into the existence of a
critical temperature in the field theory. For various reasons, it
is difficult to establish exactly what occurs at the critical
temperature in the field theory. This situation will suggest
various directions for future research.

\section{Generalised AdS black holes and a tensor perturbation}

A $D$ dimensional black hole spacetime has metric
\be
\label{eq:metric} ds^2_D = - f(r) dt^2 + \frac{dr^2}{f(r)} + r^2
d\tilde{s}^2_d ,
\ee
where $d\tilde{s}^2_d$ is a Riemannian metric on a $d=D-2$
dimensional manifold $B$, call it the horizon manifold. The black
holes will be called generalised because $B$ need not be a round
sphere. We are interested in the stability of such generalised
black hole spacetimes.

To study the stability of a solution to Einstein's equations,
consider a small transverse tracefree perturbation to the metric.
The first order change in the Ricci tensor is
\bea\label{eq:gauge}
 & & g_{ab} \to g_{ab} + h_{ab},
\quad\mbox{such that}\quad h^a{}_a
= \nabla^a h_{ab} = 0 \nonumber \\
& & R_{ab} \to R_{ab} + \frac{1}{2} (\D_L h)_{ab} ,
\eea
where the Lichnerowicz operator acting on a symmetric second rank
tensor $h$ is
\be\label{eq:lich}
(\D_L h)_{ab} = 2 R^c{}_{abd} h^d{}_c + R_{ca}
h^c{}_b + R_{cb} h^c{}_a - \nabla^c \nabla_c h_{ab} .
\ee
Gauge freedom was discussed in \cite{Gibbons:2002pq} and shall not
be of concern here.

Linearised perturbations to black hole spacetimes may be separated
into scalar, vector and tensor modes with respect to the horizon.
It was shown in \cite{Gibbons:2002pq} that the dangerous mode for
generalised black holes spacetimes is a tensor mode, which
satisfies the further conditions
\be\label{eq:extra}
h_{0a} = h_{1a} = 0 ,
\ee
where $0,1$ are the $t,r$ coordinates. The conditions
(\ref{eq:extra}) and the form of the metric (\ref{eq:metric})
imply that the transverse tracefree property of $h_{ab}$
(\ref{eq:gauge}) is inherited by $h_{\a\b}$. The indices
$a,b,\ldots$ run from $0\ldots D$ and the indices $\a,\b,\ldots$
will run from $2\ldots D$, the coordinates on $B$. Thus the mode
is a tensor mode on the $d$-dimensional horizon manifold $B$.

Unstable modes are bounded, normalisable modes
\cite{Gibbons:2002pq} that grow in time,
\be\label{eq:mode}
h_{\a\b} = \widetilde{h}_{\a\b}(x) r^2 \vp(r) e^{\w t} ,
\ee
where $x$ are coordinates on $B$ and we have decomposed the mode
into Lichnerowicz harmonics on the horizon manifold, $B$,
\be
(\widetilde{\D}_L \tilde{h})_{\a\b} = \lambda
\widetilde{h}_{\a\b} .
\ee
Tildes denote tensors on $B$. Thus $\widetilde{\D}_L$ is the
Lichnerowicz operator on $B$. Note that although the mode
(\ref{eq:mode}) only has legs along the horizon directions, it has
a radial dependence, transverse to the horizon. Radial boundary
conditions, at the horizon and at infinity, are discussed below.

We are presently interested in generalised black holes in Anti-de
Sitter spacetime. Thus the metric (\ref{eq:metric}) must solve the
vacuum Einstein equations with a negative cosmological constant
\be
R_{ab} = - H^2 (d+1) g_{ab} .
\ee
We will be considering the case when the metric on $B$ is Einstein
with positive curvature
\be\label{eq:horizon}
\tilde{R}_{\a\b}=(d-1) \tilde{g}_{\a\b} .
\ee
Again, tildes denote tensors on $B$. The normalisation corresponds
to having the same scalar curvature as $S^d$. To satisfy the
Einstein equations, incorporating (\ref{eq:horizon}), the function
$f$ must be of the form
\be\label{eq:fform}
f(r) = 1-\left(\frac{\a}{r}\right)^{d-1} + H^2 r^2.
\ee
Here $\a^{d-1}$ is proportional to the mass of the black hole.

The perturbation must satisfy the linearised equation $\d
R_{\a\b}=-H^2(d+1) h_{\a\b}$. This gives an equation for $\vp$
that may be written as a Schr\"odinger equation
\cite{Gibbons:2002pq} by changing variables to Regge-Wheeler type
coordinates and rescaling
\be\label{eq:change}
d\rs = \frac{dr}{f} ,\quad\quad \P = r^{\frac{d}{2}} \vp .
\ee
The equation for the perturbation becomes
\be\label{eq:schro}
-\frac{d^2\P}{d\rs^2} + V(r(\rs)) \P = - \w^2 \P \equiv E \P,
\ee
where the potential is
\be\label{eq:V}
V(r) = \frac{\lambda f}{r^2}
+ \frac{d-4}{2}\frac{f^{\prime}f}{r} +
\frac{d^2-10d+8}{4}\frac{f^2}{r^2} + 2H^2(d+1) f.
\ee
The stability problem reduces to the existence of bound states
with $E < 0$ of the Schr\"odinger equation with potential
$V(r(\rs))$. If such a bound state of the Schr\"odinger equation
exists, then the spacetime (\ref{eq:metric}) is unstable to modes
of the form (\ref{eq:mode}). That is to say, there will be an
instability if the ground state eigenvalue, $E_0$, of
(\ref{eq:schro}) is negative.

The mode must also satisfy boundary conditions at the horizon and
at radial infinity. Firstly, the mode must remain bounded, so that
the linearised approximation is valid. Secondly, one requires that
the mode is normalisable, $\int \P^2 d\rs < \infty$. This second
issue was shown in \cite{Gibbons:2002pq} to be equivalent to
requiring finite energy of the spacetime mode (\ref{eq:mode}). In
practice, for undesirable modes $\P(r)$ diverges at the horizon or
at infinity. These issues are discussed in some detail in
\cite{Gibbons:2002pq}. All modes considered in this work are
well-behaved.

Explicitly, for the case of an AdS black hole, the potential
becomes
\bea\label{eq:potential}
V(r) & = & \left[
\frac{d^2-10d+8+4\lambda}{4 r^2} + \frac{(d^2+2d) H^2}{4} \right]
(H^2 r^2 + 1) \nonumber \\
 & & +\frac{1}{\a^2}
\left(\frac{\a}{r}\right)^{d+1} \left[ \frac{10d-8-4\lambda-2d H^2
r^2}{4} -\left(\frac{\a}{r}\right)^{d-1} \frac{d^2}{4} \right].
\eea
It is clear that an analytic solution to the Schr\"odinger
equation is out of the question. Further, the ideas used in
\cite{Gibbons:2002pq}, which may be made rigorous, to derive a
criterion for instability for generalised flat space black holes
are not so easily applied in this case. Another argument is
needed.

\section{A criterion for instability}

Let us introduce the horizon radius $r_+$, satisfying $f(r_+)=0$.
In terms of this radius, the function (\ref{eq:fform}) becomes
\be\label{eq:rplus}
f(r) = 1 - \frac{(1+H^2 r_+^2)r^{d-1}_+}{r^{d-1}} + H^2 r^2 .
\ee

To be in a regime in which gravity is expected to be a valid low
energy theory, we need the spacetime curvature to be small outside the
horizon, which requires $H$ to be small. In
this section we will further restrict attention to large black
holes with
\be\label{eq:largeH}
H r_+ \gg 1 .
\ee
Although we are considering a classical gravitational instability
here, we will later be interested in finding a dual field theory
description. In that context, we want to be in the regime in which
the Euclidean black hole makes the dominant contribution to the
quantum gravity partition function
\cite{Hawking:1982dh,Witten:1998zw}, as opposed to periodically
identified, generalised Euclidean AdS space. The large black hole
condition (\ref{eq:largeH}) ensures that we are safely within this
regime.

More will be said about the large black hole limit below. Small
black holes, of less interest to us here, will be discussed in the
following section.

The criterion for classical dynamical instability will be a
critical value for the lowest eigenvalue of the Lichnerowicz
operator on the horizon manifold $B$ \cite{Gibbons:2002pq}. If the
eigenvalue is below this critical value, then the potential
(\ref{eq:potential}) admits a bound state with negative energy. As
we raise the eigenvalue towards the critical value, the energy of
the bound state is raised to zero. Thus to find the critical value
itself $\lambda_c$, set $\w = 0$. Note that this wouldn't have
worked for flat space black holes, because in that case the
potential remains negative at large radial direction, and there is
no zero energy bound state.

Using (\ref{eq:largeH}) in the Schr\"odinger equation
(\ref{eq:schro}) and further putting $\w=0$, one obtains another
Schr\"odinger type equation, now with a weight function,
\be\label{eq:s2}
-\frac{d^2\P}{d\rho_*^2} + \widetilde{V}(\rho(\rho_*)) \P =
\frac{-\lambda_c}{H^2 r_+^2} k(\rho(\rho_*))\P,
\ee
where we have set $\rho = r/r_+$, so that the horizon is now at
$\rho=1$, and
\bea
\frac{d\rho}{d\rho_*} & = & \rho^2 - 1/\rho^{d-1},\nonumber \\
\widetilde{V}(\rho) & = & \frac{1}{4}\left[\rho^2 - 1/\rho^{d-1}\right] \left[
2d+d^2 + d^2/\rho^{d+1} \right] ,\nonumber \\
k(\rho) & = & 1 - 1/\rho^{d+1} .
\eea
We are now looking for the lowest bound state of this new
Schr\"odinger equation (\ref{eq:s2}). This will give us
$\lambda_c$ and hence the criterion for instability. Note that the
potential is positive and further $H r_+ \gg 1$; therefore
$\lambda_c$ will be large and negative.

It seems that this equation cannot be solved analytically except
for the case when $d=1$. However, there are now no undetermined
parameters in the potential and so it is not difficult to find the
lowest eigenvalue numerically. The results for a range of
dimensions are shown in Table 1.

\begin{table}[h]
  \centering
  \caption{Critical eigenvalues for various dimensions}
\begin{tabular}{|c|c|c|c|c|c|c|} \hline
  $d$ & 1 & 2 & 3 & 4 & 5 & 6 \\ \hline
  $-\frac{\lambda_c}{H^2 r_+^2}$ & 4 & 7.4080 & 11.588 & 16.494 & 22.097 & 28.375 \\
  \hline\hline
  $d$ & 7 & 8 & 9 & 10 & 11 & 12 \\ \hline
  $-\frac{\lambda_c}{H^2 r_+^2}$ & 35.313
  & 42.897 & 51.118 & 59.966 & 69.434 & 79.515 \\ \hline
\end{tabular}
\end{table}

One may do a least squares fit to this data using a quadratic
function in $d$. To get reliable statistics, we consider the
numerically found eigenvalues for $110$ values for $d$, spaced
evenly between $d=1$ and $d=12$. The result is
\be\label{eq:Hcrit}
\lambda_c \approx - H^2 r_+^2\left[0.86 + 2.6 d + 0.33 d^2 \right] .
\ee
The fit is good in the range considered, see Figure 1 below,
although less good for the smaller values of $d$. A quadratic fit is natural because
higher powers of $d$ would have very small coefficients.

\begin{figure}[h]
  \centering
  \epsfig{file=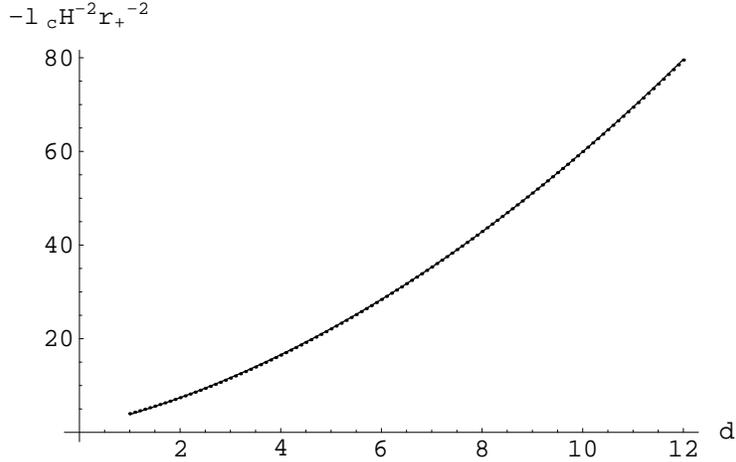,width=10cm}
  \caption{$-\frac{\lambda_c}{H^2 r_+^2}$ against $d$, datapoints
  and fitted curve shown.}
\end{figure}

The criterion for instability is thus, for $H r_+ \gg 1$, whether
the horizon manifold $B$ admits a Lichnerowicz eigenvalue lower
than the critical value (\ref{eq:Hcrit}),
\be
\lambda < \lambda_c \quad\Leftrightarrow \quad \text{Instability}.
\ee

The existence and properties of B\"ohm metrics \cite{bohm} show
that this instability criterion is not vacuous. Some properties of
these metrics are reviewed in the second half of \S 5 of this
paper. The essentials are as follows. The B\"ohm metrics are
infinite sequences of nonsingular inhomogeneous Einstein metrics
on $S^p \times S^{d-p}$ and $S^d$ for $5 \leq d \leq 9$ and $p
\geq 2$, $d-p \geq 2$. These metrics may therefore be used on
horizon manifolds. It was shown in \cite{Gibbons:2002th} that the
sequence of B\"ohm metrics on $S^2\times S^3$ and $S^5$ admit an
increasingly negative Lichnerowicz eigenvalue. For example, the
metric on $S^5$ denoted B\"ohm$(2,2)_6$ in \cite{Gibbons:2002th}
has a Lichnerowicz eigenvalue lower than $-1040$, assuming the
curvature normalisation of (\ref{eq:horizon}). Other metrics on
$S^5$ in the sequence, denoted B\"ohm$(2,2)_{2m}$, are expected to
have arbitrarily lower Lichnerowicz eigenvalues. Thus, whatever
value of $H r_+$ is needed in order for the approximation
(\ref{eq:largeH}) to be valid, there will be a B\"ohm metric with
a Lichnerowicz eigenvalue lower than the corresponding critical
value of equation (\ref{eq:Hcrit}). The generalised black hole
spacetime with this horizon manifold will then be unstable.

As we increase $m$ in the sequence of metrics B\"ohm$(2,2)_{2m}$,
the metrics, whilst remaining nonsingular, have increasingly large
curvature at two points. They tend to the singular ``double cone''
in the limit $m\to\infty$ . However, we will see in \S 5 that by
considering a double scaling limit, in which both the black hole
size and the curvature on the horizon manifold are scaled, one may
always obtain unstable B\"ohm black holes with a maximum spacetime
curvature that is well within the regime of validity for classical
gravity.

\section{Small black holes}

Normalisability of the perturbation in the infinite spacetime
volume implies that the instability is localised near the event
horizon of the black hole. Therefore, we should expect that for
small generalised black holes, $Hr_+
\ll 1$, the criterion for classical instability will just be the flat
space criterion
found in \cite{Gibbons:2002pq}. In this section we check that the
flat space instability criterion is indeed recovered. Independently of
the classical instability criterion, small black
holes are thermodynamically unstable and do not contribute
dominantly to the quantum gravity partition function
\cite{Hawking:1982dh,Witten:1998zw}.

The strategy to find the critical Lichnerowicz eigenvalue will be
to separately solve the Schr\"odinger equation (\ref{eq:schro})
with $\w=0$ in the regions $\frac{r}{r_+} \gg 1$ and
$\frac{r}{r_+} \ll \frac{1}{H r_+}$. Because $Hr_+$ is small, both
solutions will be valid in the non-empty overlap $1 \ll
\frac{r}{r_+}\ll \frac{1}{H r_+}$. A bound state will exist if the
well-behaved solution at infinity matches onto a well-behaved
solution at the horizon.

In the region $\frac{r}{r_+} \ll \frac{1}{H r_+}$, or equivalently
$H r \ll 1$, the Schr\"odinger equation (\ref{eq:schro}) reduces
to the flat space equation with $H=0$. This is immediately seen
from equations (\ref{eq:fform}) and (\ref{eq:potential}). The
solution to this equation that is regular at the horizon may be
found in terms of a hypergeometric function
\be\label{eq:nearhorizon}
\P_1(r) = (r/r_+)^{d/2} {}_2 F_1 \left(\frac{1}{2} + \frac{Q}{2(d-1)},
\frac{1}{2}-\frac{Q}{2(d-1)},1; 1- (r/r_+)^{d-1}
\right),
\ee
where $Q = \sqrt{4\lambda - 16 +(d-5)^2}$.

Now consider the region $\frac{r}{r_+} \gg 1$. In this region, the
Schr\"odinger equation (\ref{eq:schro}) becomes the equation in
pure generalised AdS, with no black hole present, e.g. with
$r_+=0$. This is easily seen from equation (\ref{eq:rplus}). The
solution to this equation that decays at infinity was written down
in \cite{Gibbons:2002pq} and is again given in terms of a
hypergeometric function
\be
\P_2(r) = r^{-(2+d)/2} {}_2 F_1 \left(\frac{3+d+Q}{4},\frac{3+d-Q}{4},\frac{6+2d}{4};
\frac{-1}{H^2 r^2} \right) ,
\ee
where as before $Q = \sqrt{4\lambda - 16 +(d-5)^2}$.

To match these two solutions in the overlapping region of
validity, require
\be
\lim_{\frac{r}{r_+} \gg 1} \P_1(r) \propto \lim_{H r \ll 1} \P_2(r) .
\ee
These limits may be calculated using standard properties of
hypergeometric functions. We find
\bea
Q \neq 0 \Rightarrow
\left\{ \begin{array}{c}
\P_1(r) \sim r^{(1+Q)/2} \quad \text{for} \quad \frac{r}{r_+} \gg 1
, \\
\P_2(r) \sim r^{(1-Q)/2} \quad \text{for} \quad H r \ll 1 ,
\end{array} \right. \nonumber \\
Q = 0 \Rightarrow
\left\{ \begin{array}{c}
\P_1(r) \sim r^{1/2} \ln r \quad \text{for} \quad \frac{r}{r_+} \gg 1
, \\
\P_2(r) \sim r^{1/2} \ln r \quad \text{for} \quad H r \ll 1 ,
\end{array} \right.
\eea
Thus, the two limits generically agree if and only if $Q=0$. Note
that $Q$ pure imaginary will generally not match because the two
solutions will not be in phase. The instability criterion is
therefore
\be\label{eq:flatcrit}
\lambda < \lambda_c = 4 - \frac{(d-5)^2}{4} \quad\Leftrightarrow
\quad \text{Instability}.
\ee
This is the same criterion as was found for flat space black holes
\cite{Gibbons:2002pq}.

Interpolating between our two results on critical Lichnerowicz
eigenvalues gives a picture for the behaviour of $\lambda_c$ as a
function of the black hole size, $H r_+$. For small black holes,
the critical eigenvalue is a constant (\ref{eq:flatcrit}). As the
black hole gets larger, the critical eigenvalue is lowered. For
large black holes the critical eigenvalue is increasingly negative
(\ref{eq:Hcrit}).

A useful way of thinking about the instability is as follows.
Suppose a black hole with a given horizon manifold is unstable in
the small black hole limit. As we vary the black hole size, there
will be a critical value such that if the black hole is smaller
than the critical size it will be unstable and if it is larger it
will be stable. Note that the Lichnerowicz spectrum of an Einstein
metric on a compact manifold admits a lower bound in terms of the
Weyl curvature tensor \cite{Page:1984ad,Gibbons:2002th}, so such a
critical size, above which the black hole is stable, always
exists.

\section{Thermodynamics and the black brane limit}

In this section it will be seen that our classically unstable large
black holes are locally thermodynamically stable. For standard AdS
black holes, where the horizon is a round sphere, the physics in
the large size limit is locally that of a black brane, in which
the horizon may be considered to be noncompact. A recent
conjecture of Gubser and Mitra  states that translationally
invariant black branes are locally thermodynamically stable if and
only if they are classically stable
\cite{Gubser:2000mm,Gubser:2000ec,Reall:2001ag}. To see whether we
have just found counterexamples to this conjecture, we need to
understand whether the large generalised black holes may be
locally understood as translationally invariant black branes.

The thermodynamics of the generalised AdS black holes we are
considering is the same as for standard AdS black holes. The large
black hole limit, $Hr_+
\gg 1$, together with the small curvature condition $H \ll 1$, implies
that $r_+ \gg 1$. In this limit the thermodynamic quantities
associated with AdS black holes, temperature (T), energy (M),
entropy (S) and heat capacity (C), are well known, e.g.
\cite{Witten:1998zw},
\bea\label{eq:thermo}
T \approx \frac{(d+1) H^2 r_+}{4\pi} , \nonumber \\
M = E \approx \frac{d \text{Vol}(B) H^2 r_+^{d+1}}{16 \pi G} ,
\nonumber
\\
S \approx \frac{\text{Vol}(B) r_+^d}{4G} , \nonumber \\
C = \frac{dE}{dT} \approx \frac{d \text{Vol}(B) r_+^d}{4 G} \approx dS .
\eea
The last of these quantities shows that we are comfortably in the
regime where the heat capacity is positive and hence the black
hole is locally thermodynamically stable.

For the standard AdS black holes with spherical horizons, the
large black hole limit is locally the black brane limit.
Qualitatively, this is because all curvatures at the horizon
are negligible in the large black hole limit and the sphere, $S^d$,
may be locally considered to be noncompact flat space, $\R^d$.

The situation for generalised AdS black holes is a little more
subtle. In the limit where $H r_+
\gg 1$ the black hole metric may be written as
\bea\label{eq:brane}
ds^2 & = & \r^2 \left[-g(\r) H^2 r_+^2 dt^2 + r_+^2 d\tilde{s}^2_d
\right] + \frac{d\r^2}{\r^2 H^2 g(\r)},
\nonumber \\
g(\r) & = & 1 - \r^{-d-1},
\eea
where $\r = r/r_+$, so the horizon is at $\r=1$. If $B=S^d$ one
can now locally replace $r_+^2 d\tilde{s}^2_d$ by the flat metric
$dx_1^2 + \cdots + dx_d^2$, because $H$ is small so $r_+$ is large.
The resulting spacetime (\ref{eq:brane}) then describes a black
brane with $d$ noncompact, translationally invariant, spatial
directions and one transverse direction. We will call $r_+^2
d\tilde{s}^2_d$ the rescaled horizon metric. It is the metric of the
horizon embedded into the spacetime.

The first point we can make in the generalised case is that whilst
the scalar curvature of $B$ rescaled by $d\tilde{s}^2_d \to r_+^2
d\tilde{s}^2_d$ is small, going as $\tilde{R} \to d(d-1)/r_+^2$,
we will now construct another curvature invariant which is not
vanishingly small for large $r_+$.

For compact manifolds with Einstein metrics, the Lichnerowicz
spectrum on transverse tracefree modes is effectively bounded
below by the Weyl tensor. Let $\lambda$ be the minimum
Lichnerowicz eigenvalue. Let $\kappa(x)$ be the, position
dependent, eigenvalue of the Weyl tensor
\be
\tilde{C}_{\a\b}{}^{\g\d}(x) \tilde{h}_{\g\d}(x)
= \kappa(x) \tilde{h}_{\a\b}(x),
\ee
such that $\kappa_0 \equiv \parallel \kappa(x) \parallel_{\infty}
= \sup_{x \in B} \kappa(x)$ is maximised. Think of this as the
largest eigenvalue of the Weyl tensor. Then one has
\cite{Gibbons:2002th}
\be\label{eq:weylbound}
\lambda \geq 4d - 4 \kappa_0 .
\ee
For the unstable horizon manifolds, we have seen that $\lambda$ is
large and negative and therefore $\kappa_0$ is large and positive
and the $4d$ term is negligible. Further, we found above in
(\ref{eq:Hcrit}) that $\mid \lambda \mid \gtrsim H^2 r_+^2$.
Putting these results together allows us to bound a curvature
invariant
\be\label{eq:C0bound}
C_0 \equiv \;\parallel \tilde{C}_{\a\b\g\d} \tilde{C}^{\a\b\g\d}
\parallel_{\infty}
\;\geq\;
\kappa_{0}^2
\;\geq \frac{\lambda^2}{16}\;
\gtrsim H^4 r_+^4 .
\ee
Now, under the rescaling $d\tilde{s}^2_d \to r_+^2
d\tilde{s}^2_d$, one has $\tilde{C}_{\a\b\g\d}
\tilde{C}^{\a\b\g\d}
\to r_+^{-4} \tilde{C}_{\a\b\g\d} \tilde{C}^{\a\b\g\d}$. But this now
implies that $C_0 \to r_+^{-4} C_0 \sim H^4$, which is finite!
This behaviour should be contrasted with the behaviour of the
rescaled scalar curvature which becomes small, as noted above.

Two short comments are appropriate. Firstly, note that the
rescaled $\tilde{C}_{\a\b\g\d} \tilde{C}^{\a\b\g\d}$ is of order
$H^4$ which, although finite, is small. It is easy to check that
this implies that curvatures of the full black hole solution are
small outside the horizon, if $r_+$ is large, and therefore we are
within the regime of validity of the supergravity description.
Secondly, we are effectively considering a double scaling limit.
Both the horizon radius and the absolute value of the Lichnerowicz
eigenvalue are large, but the relationship (\ref{eq:weylbound})
essentially tells us that the limit is taken in such a way that
$\tilde{C}_{\a\b\g\d} \tilde{C}^{\a\b\g\d}$ remains finite at the
horizon.

The persistence of a finite curvature invariant as the horizon
manifold is rescaled suggests that a translationally invariant black
brane picture is problematic. To see to what
extent this is the case, consider the example of B\"ohm metrics,
mentioned at the end of \S 3.

The rescaled horizon B\"ohm metrics \cite{bohm,Gibbons:2002th} are
of the form
\be\label{eq:bohmmetric}
r_+^2 d\tilde{s}^2_d = r_+^2 \left[ d\q^2 + a(\q)^2 d\Omega^2_p
+ b(\q)^2 d\widetilde{\Omega}^2_q \right],
\ee
where $d\Omega^2_p$ and $d\widetilde{\Omega}^2_q$ are round
metrics on $S^p$ and $\tilde{S}^q$, respectively. Clearly
$p+q+1=d$. The functions $a(\q),b(\q)$ are determined by the
Einstein condition. Let us restrict attention to the cases where
$p=q=2$ and the topology is $S^3\times S^2$. Other cases are
essentially the same, including when the topology is that of a
sphere. In the $S^3\times S^2$ case, $a(\q)$ vanishes at $\q=0$
and $\q=\q_{\text{max}}$, and $b(\q)$ does not vanish. Both $a$
and $b$ are symmetric about the midpoint $\q_{\text{max}}/2$.

The finite curvature, when $r_+$ is large, is concentrated around
$\q=0,\q_{\text{max}}$. Near $\q=0$ we have \cite{Gibbons:2002th}
\be
a(\q) = \q - \frac{2 b_0^2+1}{18 b_0^2} \q^3 + \cdots \quad ;
\quad b(\q) = b_0 - \frac{4 b_0^2-1}{6 b_0} \q^2 + \cdots .
\ee
There is a sequence of such metrics, given by a sequence of allowed
values for $b_0$. The first is the standard
metric on $S^3\times S^2$. For the other metrics, $b_0$ becomes
increasingly small. Thus, if we think of (\ref{eq:bohmmetric}) as
$\tilde{S}^2$ fibred over $S^3$ then as $b_0 \to 0$ the base $S^3$
starts to look like a higher dimensional rugby ball (or American
football) and the fibre $\tilde{S}^2$ is small at the endpoints
$\q=0,\q_{\text{max}}$. In the limit $b_0 = 0$ one obtains a
singular ``double-cone'' over $S^2\times S^2$. This may be
quantified by calculating the rescaled curvature invariants
\bea
\left. \tilde{R}_{\a\b\g\d} \tilde{R}^{\a\b\g\d}\right|_{\text{rescaled}} & = & \frac{8 - 16 b_0^2 + 48
b_0^4}{r_+^4 b_0^4}\quad\text{at}\quad \q = 0, \nonumber \\
\left. \tilde{C}_{\a\b\g\d} \tilde{C}^{\a\b\g\d}\right|_{\text{rescaled}} & = & \frac{8 - 16 b_0^2 + 8
b_0^4}{r_+^4 b_0^4}\quad\text{at}\quad \q = 0 ,
\eea
These are in fact the maximum values obtained by the curvatures on
the manifold. We have just seen that instability requires, for the
rescaled curvatures, $\tilde{C}_{\a\b\g\d} \tilde{C}^{\a\b\g\d} \sim H^4$, and
therefore $b_0 \sim 1/(H r_+)$.

In contrast to the behaviour at the endpoints, if we expand $a$
and $b$ about the midpoint $\q_{\text{max}}/2$ and calculate the
rescaled curvatures we find
\be
\left. \tilde{R}_{\a\b\g\d} \tilde{R}^{\a\b\g\d} \right|_{\text{rescaled}} \sim
\left. \tilde{C}_{\a\b\g\d} \tilde{C}^{\a\b\g\d} \right|_{\text{rescaled}} \sim
\frac{{\mathcal{O}}(1)}{r_+^4}\quad\text{at}\quad \q = \frac{\q_{\text{max}}}{2},
\ee
which is small for $r_+ \gg 1$. Because we are working with a
Riemannian metric, $\tilde{R}_{\a\b\g\d} \tilde{R}^{\a\b\g\d} \to
0$ is equivalent to $\tilde{R}_{\a\b\g\d} \to 0$, which then
implies that the space is becoming flat.

Therefore, if we are located away from the endpoints of the B\"ohm
metrics, the large black hole limit does correspond locally to a
black brane limit with a flat, translationally invariant, horizon.
However, if we are located very near the endpoints, then a finite
curvature persists in the large black hole limit, and the horizon
may not be locally approximated by a metric with noncompact,
translationally invariant directions.

Given these observations, it would be misleading to describe the
large black hole limit as a black brane limit, although it is a
locally accurate description away from the endpoints. Furthermore,
the precise ``ballooning'' instability of the B\"ohm metrics
\cite{Gibbons:2002th}, in which one of the $S^2$s grows and the
other shrinks, crucially requires the finite curvature. This
should not be surprising given that when there is no curvature,
there is no instability. Thus, the strict statement of the
Gubser-Mitra conjecture, requiring translational invariance,
remains intact.

However, the results presented here suggest that one should be
cautious about the general validity of using local thermodynamic
arguments to understand classical instabilities.

\section{Field theory implications of the instability}

The generalised black holes we have been studying have a dual
description using the logic of the AdS/CFT correspondence.
A generalised AdS black hole with horizon manifold $B$ will be dual to
a thermal field theory on $S^1\times B^{\prime}$
\cite{Witten:1998zw,Birmingham:1998nr}, where $B^{\prime}$ is a scaled
up copy of $B$, to be specified shortly. We should be more precise
about the size of the $S^1$ and $B^{\prime}$, that is, the dual temperature and
spatial volume.

The temperature of the dual field theory at cutoff radius $r=r_0$ is given by
the local, redshifted, temperature of the black hole. For $r_0 \gg
r_+$ one has in the large black hole limit, using (\ref{eq:thermo}),
\be
T_{\text{FT}} = \frac{T}{\sqrt{-g_{tt}}} \approx \frac{T}{H r_0} \approx \frac{(d+1)}{4\pi r_0} H r_+ .
\ee
The large black hole limit is thus the high temperature limit.

The volume of the spatial section at $r=r_0$ is just
$\text{Vol}(B^{\prime}) = r_0^d \text{Vol}(B)$. Thus metrically we have
$ds^2_{B^{\prime}} = r_0^2 ds^2_B$.

The instability criterion of (\ref{eq:Hcrit}) may be rewritten in
terms of the field theory temperature. The
solution is unstable if
\be\label{eq:fieldtheory}
- \lambda^{\prime} \; = \; - \frac{\lambda}{r_0^2} \; \gtrsim \;
T_{\text{FT}}^2 \frac{16 \pi^2 (0.86 + 2.6 d + 0.33 d^2)}{(d+1)^2} ,
\ee
where $\lambda^{\prime}$ is the Lichnerowicz eigenvalue on
$B^{\prime}$ corresponding to $\lambda$ on $B$. Note that the final
expression contains no explicit reference to the cutoff radius
$r_0$. The criterion is expressed purely in terms of the Lichnerowicz
spectrum of a manifold $B^{\prime}$ and the temperature of a field
theory on that manifold.

Therefore, the duality predicts some effect in the thermal field
theory on $B^{\prime}$ when the temperature drops below a certain
value. The effect could be called a finite curvature effect because
the previously discussed relationship between $\lambda$ and
the Weyl squared curvature invariant (\ref{eq:C0bound}) implies that when
(\ref{eq:fieldtheory}) is satisfied, the temperature is
small compared to the curvature of $B^{\prime}$ in certain regions.

The precise nature of the effect remains elusive for several
reasons. Firstly, B\"ohm metrics currently provide the only known
horizon manifolds that give unstable large AdS black holes, and
only exist in dimensions $d=5..9$. Therefore, the immediate
application to a known duality is the $AdS_7 \times S^4$ version
of the AdS/CFT correspondence. However, the field theory dual is
not under computational control in this case. The question of
finding unstable large black holes in lower dimensional generalised
AdS spaces, $AdS_5$ and $AdS_6$, depends on the existence of suitable
positive curvature Einstein manifolds in three and four
dimensions, respectively. In three dimensions, positive curvature
Einstein manifolds are quotients of the round sphere \cite{besse}
and therefore cannot have negative Lichnerowicz eigenvalues. In
four dimensions, the situation is not clear, although some
rigidity results exist \cite{besse,Yang}.

Secondly, given that the unstable generalised black holes cannot
provide the dual description to field theory physics below the
critical temperature, new stable gravity solutions are needed. It is
not known whether such solutions, stable and with appropriate $S^1
\times B^{\prime}$ asymptotics, exist or what properties they might
have. Understanding whether these stable solutions exist could also
shed some light upon the endpoint of the classical instability.

There are also various issues concerning the thermal field theory which
one would like to understand better. Is the effect specific to
exotic field theories such as that dual to M-theory on $AdS_7
\times S^4$ or is it generic? Does the spatial background
$B^{\prime}$ need to be Einstein, or is it sufficient to have a
Lichnerowicz spectrum satisfying (\ref{eq:fieldtheory})?

\section*{Acknowledgments}

During this work I have had stimulating conversations with David
Berman, Christophe Patricot, Guiseppe Policastro, Rub\'{e}n
Portugu\'{e}s, Guillermo Silva, Nemani Suryanarayana and Toby
Wiseman. I'd also like to thank Gary Gibbons for comments on the
text. The author is funded by the Sims scholarship.

\end{document}